\documentstyle[12pt]{article}
\begin{document}

\author{C. Bizdadea\thanks{e-mail address: 
bizdadea@central.ucv.ro}\\
Department of Physics, University of Craiova \\
13 A. I. Cuza Str., Craiova RO-1100, Romania}
\title{On the cohomological derivation of 
topological Yang-Mills theory}
\maketitle

\begin{abstract}
Topological Yang-Mills theory is derived in 
the framework of Lagrangian BRST
cohomology.

\end{abstract}

\section{Introduction}

The cohomological understanding of the 
BRST symmetry \cite{1}--\cite{8}
allowed, apart from proving the 
existence of the BRST generator for an
arbitrary gauge system \cite{8}--\cite{9}, 
a useful investigation of many
interesting aspects related to 
perturbative renormalization problem \cite{10}%
--\cite{12}, anomaly-tracking 
mechanism \cite{12}--\cite{13}, simultaneous
study of local and rigid invariances 
of a given theory \cite{14}, as well as
to the construction of consistent 
interactions in gauge theories \cite{15}--%
\cite{18}. The last topic is 
probably the most efficient proof of the power
of cohomological BRST ideas, 
reformulating the classical Lagrangian problem
of building consistent interactions 
in gauge theories in terms of precise
cohomological classes of the 
BRST differential, which further offers a
systematic search for all 
possible consistent interactions in the natural
background of the deformation theory. 
Among the models of great interest in
theoretical physics that have been 
inferred along the deformation of the
master equation we mention Yang-Mills 
theory \cite{19}, the
Freedman-Townsend model \cite{20}, 
the Chapline-Manton model \cite{21}.
Also, it is important to notice the 
deformation results connected to
Einstein's gravity theory \cite{22}, 
four- and eleven-dimensional
supergravity \cite{23}, $p$-forms 
\cite{24} or chiral forms \cite{25}.
However, there remain some important 
coupled models that have not been
recovered so far in the light of the 
deformation of the master equation,
like the topological Yang-Mills theory. 
This is precisely the main aim of
our paper, namely, to infer the 
four-dimensional topological coupling among
Yang-Mills fields via the deformation technique. 
In view of this, we begin
with a certain uncoupled theory in four 
dimensions and derive its associated
BRST symmetry. Consequently, we write 
down the equations that should be
satisfied by the deformed solution to 
the master equation in terms of the
coupling constant, and find their 
consistent solutions by using the BRST
symmetry for the free model. In this manner, 
we find a complete deformed
solution, which is consistent at all 
orders in the coupling constant. From
the analysis of the structure of 
this solution we observe that the resulting
coupled model is nothing but the 
topological Yang-Mills theory. Thus, the
procedure applied in our paper 
leads to a nice example of simultaneous
deformation of the gauge transformations, 
gauge algebra and reducibility
relations of the starting uncoupled system.

\section{BRST symmetry for the uncoupled model}

Initially, we infer the antifield-BRST 
symmetry for an uncoupled model,
described by the Lagrangian action 
\begin{equation}
S_{0}^{L}\left[ A_{\mu }^{a}\right] =
-\frac{1}{4}
\int d^{4}x\varepsilon _{\mu \nu \lambda \rho }
F_{a}^{\mu \nu }F^{a\lambda \rho }.  \label{1}
\end{equation}
The field strength is defined by 
$F_{a}^{\mu \nu }=\partial ^{\mu }A_{a}^{\nu }-
\partial ^{\nu }
A_{a}^{\mu }\equiv \partial _{{}}^{\left[ \mu \right. }
A_{a}^{\left. \nu \right] }$, 
while $\varepsilon _{\mu \nu \lambda \rho }$ 
denotes the completely 
antisymmetric four-dimensional symbol. Action
(\ref{1}) is invariant under the gauge transformations 
\begin{equation}
\delta _{\epsilon }\Phi ^{\alpha _{0}}=
Z_{\;\;\alpha _{1}}^{\alpha _{0}}\epsilon ^{\alpha _{1}}
\rightarrow \delta _{\epsilon }A_{\mu }^{a}=
\partial _{\mu }\epsilon ^{a}+
\epsilon _{\mu }^{a},  \label{2}
\end{equation}
with 
\begin{equation}
\Phi ^{\alpha _{0}}
\rightarrow A_{\mu }^{a},\;\epsilon ^{\alpha _{1}}
\rightarrow \left( 
\begin{array}{l}
\epsilon ^{b} \\ 
\epsilon _{\nu }^{b}
\end{array}
\right) ,\;Z_{\;\;\alpha _{1}}^{\alpha _{0}}
\rightarrow \left( \delta
_{\;\;b}^{a}\partial _{\mu },\;
\delta _{\;\;b}^{a}\delta _{\;\;\mu }^{\nu
}\right) ,  \label{2a}
\end{equation}
which are first-stage reducible. 
Indeed, if we take $\epsilon ^{a}=
\theta ^{a},$ $\epsilon _{\mu }^{a}=
-\partial _{\mu }\theta ^{a}$, 
then the gauge
transformations (\ref{2}) vanish 
identically, $\delta _{\epsilon }A_{\mu }^{a}=
0$. Consequently, the 
reducibility relations can be written like 
\begin{equation}
Z_{\;\;\alpha _{1}}^{\alpha _{0}}
Z_{\;\;\alpha _{2}}^{\alpha _{1}}=0,
\label{2b}
\end{equation}
with the first-stage reducibility matrix given by 
\begin{equation}
Z_{\;\;\alpha _{2}}^{\alpha _{1}}\rightarrow \left( 
\begin{array}{c}
\delta _{\;\;c}^{b} \\ 
-\delta _{\;\;c}^{b}\partial _{\nu }
\end{array}
\right) .  \label{3}
\end{equation}
Accordingly, the solution to the master 
equation of the uncoupled model is
expressed by 
\begin{equation}
S=S_{0}^{L}\left[ A_{\mu }^{a}\right] +
\int d^{4}x\left( A_{a}^{*\mu }\left(
\partial _{\mu }C^{a}+C_{\mu }^{a}\right) +
C_{a}^{*}\eta ^{a}-C_{a}^{*\mu
}\partial _{\mu }\eta ^{a}\right) ,  \label{4}
\end{equation}
where $C^{a}$ and $C_{\mu }^{a}$ 
stand for the fermionic ghost number one
ghosts, and $\eta ^{a}$ denote the 
bosonic ghost number two ghosts required
by the reducibility. The star 
variables $A_{a}^{*\mu }$, $C_{a}^{*}$, $%
C_{a}^{*\mu }$ and $\eta _{a}^{*}$ 
represent the antifields of the
corresponding fields/ghosts. 
The antifields $A_{a}^{*\mu }$ are fermionic
with ghost number minus one, 
$C_{a}^{*}$ and $C_{a}^{*\mu }$ are bosonic
with ghost number minus two, 
while $\eta _{a}^{*}$ are fermionic and display
ghost number minus three. 
The ghost number is defined in the standard manner
like the difference between 
pure ghost number (${\rm pgh}$) and antighost
number (${\rm antigh}$), with 
\begin{equation}
{\rm pgh}\left( A_{\mu }^{a}\right) =
{\rm pgh}\left( A_{a}^{*\mu }\right) =%
{\rm pgh}\left( C_{a}^{*}\right) =
{\rm pgh}\left( C_{a}^{*\mu }\right) ={\rm %
pgh}\left( \eta _{a}^{*}\right) =
0,  \label{5}
\end{equation}
\begin{equation}
{\rm pgh}\left( C^{a}\right) =
{\rm pgh}\left( C_{\mu }^{a}\right) =1,\;{\rm %
pgh}\left( \eta ^{a}\right) =2,  \label{6}
\end{equation}
\begin{equation}
{\rm antigh}\left( A_{\mu }^{a}\right) =
{\rm antigh}\left( C^{a}\right) =%
{\rm antigh}\left( C_{\mu }^{a}\right) =
{\rm antigh}\left( \eta ^{a}\right) =
0,  \label{7}
\end{equation}
\begin{equation}
{\rm antigh}\left( A_{a}^{*\mu }\right) =
1,\;{\rm antigh}\left( C_{a}^{*\mu
}\right) ={\rm antigh}\left( C_{a}^{*}\right) =
2,\;{\rm antigh}\left( \eta
_{a}^{*}\right) =3.  \label{8}
\end{equation}
The antifield BRST differential $s\bullet =
\left( \bullet ,S\right) $ of the
free model splits as 
\begin{equation}
s=\delta +\gamma ,  \label{10}
\end{equation}
where $\delta $ is the Koszul-Tate 
differential, and $\gamma $ denotes the
longitudinal exterior derivative 
along the gauge orbits. The symbol $\left(
,\right) $ signifies the 
antibracket in the antifield formalism.
Consequently, we find that 
\begin{equation}
\delta A_{\mu }^{a}=0,\;\gamma A_{\mu }^{a}=
\partial _{\mu }C^{a}+C_{\mu
}^{a},  \label{11}
\end{equation}
\begin{equation}
\delta C^{a}=0,\;\gamma C^{a}=
\eta ^{a},  \label{12}
\end{equation}
\begin{equation}
\delta C_{\mu }^{a}=0,\;\gamma C_{\mu }^{a}=
-\partial _{\mu }\eta ^{a},
\label{13}
\end{equation}
\begin{equation}
\delta \eta ^{a}=0,\;\gamma \eta ^{a}=
0,  \label{14}
\end{equation}
\begin{equation}
\delta A_{a}^{*\mu }=0,\;\gamma A_{a}^{*\mu }=
0,  \label{15}
\end{equation}
\begin{equation}
\delta C_{a}^{*}=
-\partial _{\mu }A_{a}^{*\mu },\;\gamma C_{a}^{*}=0,
\label{16}
\end{equation}
\begin{equation}
\delta C_{a}^{*\mu }=
A_{a}^{*\mu },\;\gamma C_{a}^{*\mu }=
0,  \label{17}
\end{equation}
\begin{equation}
\delta \eta _{a}^{*}=-\left( C_{a}^{*}+
\partial _{\mu }C_{a}^{*\mu }\right)
,\;\gamma \eta _{a}^{*}=0.  \label{18}
\end{equation}
The above formulas will be employed in 
the next section at the deformation
of the solution (\ref{4}) in a cohomological context.

\section{Deformation procedure}

A consistent deformation of action 
(\ref{1}) and of its gauge invariances
defines a deformation of the solution 
to the master equation that preserves
both the master equation and the 
field/antifield spectra \cite{15}. This
means that if 
\begin{equation}
S_{0}^{L}\left[ A_{\mu }^{a}\right] +
g\int d^{4}x\alpha _{0}+g^{2}\int
d^{4}x\beta _{0}+
O\left( g^{3}\right) ,  \label{19}
\end{equation}
is a consistent deformation of 
action (\ref{1}), with deformed gauge
transformations 
\begin{equation}
\bar{\delta}_{\epsilon }A_{\mu }^{a}=
\partial _{\mu }\epsilon ^{a}+\epsilon
_{\mu }^{a}+g\lambda _{\mu }^{a}+
O\left( g^{2}\right) ,  \label{20}
\end{equation}
then the deformed solution to the 
master equation 
\begin{equation}
\bar{S}=S+g\int d^{4}x\alpha +
g^{2}\int d^{4}x\beta +O\left( g^{3}\right)
=S+gS_{1}+g^{2}S_{2}+
O\left( g^{3}\right) ,  \label{21}
\end{equation}
should satisfy 
\begin{equation}
\left( \bar{S},\bar{S}\right) =
0,  \label{22}
\end{equation}
where 
\begin{equation}
\alpha =\alpha _{0}+
A_{a}^{*\mu }\bar{\lambda}_{\mu }^{a}+{\rm ``more".}
\label{23}
\end{equation}
The master equation (\ref{22}) 
splits according to the deformation parameter 
$g$ as 
\begin{equation}
s\alpha =\partial _{\mu }j^{\mu },  \label{24}
\end{equation}
\begin{equation}
s\beta +\frac{1}{2}\omega =
\partial _{\mu }\theta ^{\mu },  \label{25}
\end{equation}
\[
\vdots 
\]
for some local $j^{\mu }$ and $\theta ^{\mu }$, with 
\begin{equation}
\left( S_{1},S_{1}\right) =
\int d^{4}x\omega .  \label{26}
\end{equation}
We omitted the zeroth order equation 
in the coupling constant corresponding
to the (\ref{22}) as this is automatically 
verified. From (\ref{24}) we read
that the first-order non-trivial 
consistent deformations belong to $%
H^{0}\left( s|d\right) $, where $d$ 
is the exterior space-time derivative.
In the situation where $\alpha $ is 
a coboundary modulo $d$ ($\alpha
=s\lambda +\partial _{\mu }\pi ^{\mu }$), 
the corresponding deformation is
trivial (it can be eliminated by a 
redefinition of the fields).

In order to solve equation (\ref{24}), 
we expand $\alpha $ accordingly the
antighost number 
\begin{equation}
\alpha =\alpha _{0}+\alpha _{1}+\cdots +
\alpha _{I},\;{\rm antigh}\left(
\alpha _{K}\right) =K,  \label{27}
\end{equation}
where the last term in (\ref{27}) can 
be assumed to be annihilated by $%
\gamma $. Since ${\rm antigh}\left( \alpha _{I}\right) =
I$ and ${\rm gh}%
\left( \alpha _{I}\right) =0$, 
it follows that ${\rm pgh}\left( \alpha
_{I}\right) =I$. Therefore, we can 
represent $\alpha _{I}$ under the form 
\begin{equation}
\alpha _{I}=\mu _{a_{1}
\cdots a_{M}b_{1\cdots }b_{N}}\eta ^{a_{1}}\cdots
\eta ^{a_{M}}\rho ^{b_{N}}
\cdots \rho ^{b_{1}},  \label{29}
\end{equation}
where $N$ and $M$ are some nonnegative 
integers satisfying $3N+2M=I$, and $%
\mu _{a_{1}\cdots a_{M}b_{1\cdots }b_{N}}$ 
stand for some $\gamma $%
-invariant functions with 
${\rm antigh}\left( \mu _{a_{1}\cdots
a_{N}b_{1\cdots }b_{M}}\right) =
3N+2M$. In (\ref{27}), we used the notation 
\begin{equation}
\rho ^{a}=
f_{\;\;bc}^{a}\eta ^{b}C^{c},  \label{28}
\end{equation}
with $f_{\;\;bc}^{a}$ some constants 
that are antisymmetric in the lower
indices, $f_{\;\;bc}^{a}=-f_{\;\;cb}^{a}$. 
The antisymmetry of $%
f_{\;\;bc}^{a}$ is implied by the 
$\gamma $-invariance of $\rho ^{a}$. Thus,
the general form of $\mu _{a_{1}
\cdots a_{M}b_{1\cdots }b_{N}}$ is given by 
\begin{equation}
\mu _{a_{1}\cdots a_{M}b_{1\cdots }b_{N}}=
\left( -\right) ^{N}\eta
_{b_{1}}^{*}\cdots \eta _{b_{N}}^{*}\left(
\sum\limits_{k=0}^{M}C_{a_{1}}^{*}
\cdots C_{a_{M-k}}^{*}\left( \partial
_{\mu _{1}}C_{a_{M-k+1}}^%
{*\mu _{1}}\right) \cdots \left( \partial _{\mu
_{k}}C_{a_{M}}^{*\mu _{k}}\right) \right) ,  \label{30}
\end{equation}
so 
\begin{equation}
\alpha _{I}=\rho ^{N}\sum\limits_{k=0}^{M}
\lambda ^{M-k}\sigma ^{k},
\label{32}
\end{equation}
where 
\begin{equation}
\rho =-\eta _{a}^{*}\rho ^{a},\;\lambda =
C_{a}^{*}\eta ^{a},\;\sigma =\left(
\partial _{\mu }
C_{a}^{*\mu }\right) \eta ^{a}.  \label{33}
\end{equation}
Taking into account the formulas 
(\ref{11})--(\ref{18}) and the above form
of $\alpha _{I}$, we obtain 
\begin{eqnarray}
&&\delta \alpha _{I}=
\gamma \left( \left( \rho
^{N}\sum\limits_{k=0}^{M}k\lambda ^{M-k}
\sigma ^{k-1}-\left( M-k\right)
\lambda ^{M-k-1}
\sigma ^{k}\right) A_{b}^{*\mu }C_{\mu }^{b}+\right. 
\nonumber \\ \label{34}
&&\left. N\rho ^{N-1}\left( \sum\limits_{k=0}^{M}
\lambda ^{M-k}\sigma
^{k}\right) \left( \frac{1}{2}C_{a}^{*}
f_{\;\;bc}^{a}C^{b}C^{c}+C_{a}^{*\mu
}f_{\;\;bc}^{a}\left( C^{b}C_{\mu }^{c}+
\eta ^{b}A_{\mu }^{c}\right) \right)
\right) +\nonumber \\
&&\partial _{\mu }\left( \left( \rho ^{N}
\sum\limits_{k=0}^{M}k\lambda
^{M-k}\sigma ^{k-1}-
\left( M-k\right) \lambda ^{M-k-1}\sigma ^{k}\right)
A_{b}^{*\mu }\eta ^{b}+\right. \nonumber \\
&&\left. N\rho ^{N-1}\left( \sum\limits_{k=0}^{M}
\lambda ^{M-k}\sigma
^{k}\right) C_{a}^{*\mu }
f_{\;\;bc}^{a}C^{b}\eta ^{c}\right) -
\nonumber \\
&&A_{b}^{*\mu }\eta ^{b}
\partial _{\mu }\left( \rho
^{N}\sum\limits_{k=0}^{M}k\lambda ^{M-k}
\sigma ^{k-1}-\left( M-k\right)
\lambda ^{M-k-1}\sigma ^{k}\right) -
\nonumber \\
&&C_{a}^{*\mu }f_{\;\;bc}^{a}C^{b}
\eta ^{c}\partial _{\mu }\left( N\rho
^{N-1}\sum\limits_{k=0}^{M}
\lambda ^{M-k}\sigma ^{k}\right) .  
\end{eqnarray}
On the other hand, equation (\ref{24}) 
at antighost number $\left(
I-1\right) $ takes the form 
\begin{equation}
\delta \alpha _{I}+\gamma \alpha _{I-1}=
\partial _{\mu }m^{\mu },  \label{35}
\end{equation}
for some $m^{\mu }$. The equations 
(\ref{34}) and (\ref{35}) have to be
compatible. This happens if and only 
if $M=0$ and $N=1$, such that $I=3$. In
this way, we inferred that the last 
term in (\ref{27}) is given precisely by 
\begin{equation}
\alpha _{3}=\rho =-\eta _{a}^{*}
f_{\;\;bc}^{a}\eta ^{b}C^{c}.  \label{36}
\end{equation}
Now, from (\ref{34}) restricted to 
$M=0$ and $N=1$, we find 
\begin{eqnarray}
&&\delta \alpha _{3}=\gamma 
\left( \frac{1}{2}C_{a}^{*}f_{\;%
\;bc}^{a}C^{b}C^{c}+C_{a}^{*\mu }
f_{\;\;bc}^{a}\left( C^{b}C_{\mu }^{c}+\eta
^{b}A_{\mu }^{c}\right) \right) +
\nonumber \\ \label{37}
&&\partial _{\mu }\left( C_{a}^{*\mu }
f_{\;\;bc}^{a}C^{b}\eta ^{c}\right) ,
\end{eqnarray}
such that 
\begin{equation}
\alpha _{2}=-\frac{1}{2}C_{a}^{*}
f_{\;\;bc}^{a}C^{b}C^{c}-C_{a}^{*\mu
}f_{\;\;bc}^{a}\left( C^{b}C_{\mu }^{c}+
\eta ^{b}A_{\mu }^{c}\right) .
\label{38}
\end{equation}
With $\alpha _{2}$ at hand, we determine 
$\alpha _{1}$ as solution to the
equation 
\begin{equation}
\delta \alpha _{2}+\gamma \alpha _{1}=
\partial _{\mu }n^{\mu }.  \label{39}
\end{equation}
On behalf of (\ref{38}), we get 
\begin{equation}
\delta \alpha _{2}=\frac{1}{2}
\left( \partial _{\mu }A_{a}^{*\mu }\right)
f_{\;\;bc}^{a}C^{b}C^{c}-A_{a}^{*\mu }
f_{\;\;bc}^{a}\left( C^{b}C_{\mu
}^{c}+\eta ^{b}A_{\mu }^{c}\right) .  \label{40}
\end{equation}
Then, the solution to (\ref{39}) 
is expressed by 
\begin{equation}
\alpha _{1}=A_{a}^{*\mu }f_{\;\;bc}^{a}
A_{\mu }^{c}C^{b},  \label{41}
\end{equation}
which yields to 
\begin{equation}
\delta \alpha _{2}+\gamma \alpha _{1}=
\partial _{\mu }\left( \frac{1}{2}%
A_{a}^{*\mu }f_{\;\;bc}^{a}
C^{b}C^{c}\right) .  \label{42}
\end{equation}
By projecting (\ref{24}) on antighost 
number one, we deduce the relation 
\begin{equation}
\delta \alpha _{1}+\gamma \alpha _{0}=
\partial _{\mu }v^{\mu }.  \label{43}
\end{equation}
In the meantime, as $\delta \alpha _{1}=
0$, we arrive at 
\begin{equation}
\alpha _{0}=\frac{1}{2}
\varepsilon _{\mu \nu \lambda \rho
}f_{\;\;bc}^{a}F_{a}^{\mu \nu }
A^{\lambda b}A^{\rho c},  \label{44}
\end{equation}
that further leads to 
\begin{equation}
\gamma \alpha _{0}=\partial _{\mu }
\left( \varepsilon ^{\mu \nu \lambda \rho
}f_{\;\;bc}^{a}\left( C_{\rho a}
A_{\nu }^{b}A_{\lambda }^{c}+C_{a}\partial
_{\rho }\left( A_{\nu }^{b}
A_{\lambda }^{c}\right) \right) \right) .
\label{45}
\end{equation}
Putting together the results 
expressed by (\ref{36}), (\ref{38}), (\ref{41})
and (\ref{44}), it results that 
the complete first-order deformation reads
as
\begin{eqnarray}
&&S_{1}=\int d^{4}x\left( \frac{1}{2}
\varepsilon _{\mu \nu \lambda \rho
}f_{\;\;bc}^{a}F_{a}^{\mu \nu }
A^{\lambda b}A^{\rho c}+A_{a}^{*\mu
}f_{\;\;bc}^{a}A_{\mu }^{c}C^{b}-
\right. \nonumber \\ \label{46}
&&\left. \frac{1}{2}C_{a}^{*}
f_{\;\;bc}^{a}C^{b}C^{c}-C_{a}^{*\mu
}f_{\;\;bc}^{a}\left( C^{b}C_{\mu }^{c}+
\eta ^{b}A_{\mu }^{c}\right) -\eta
_{a}^{*}f_{\;\;bc}^{a}\eta ^{b}C^{c}\right) . 
\end{eqnarray}

Until now we proved the existence of 
$\alpha $ as solution to the equation (%
\ref{24}), which is equivalent with 
the consistency of the interaction to
order $g$. The interaction is also 
consistent to order $g^{2}$ if and only
if equation (\ref{25}) possesses 
solution (with respect to $\beta $), hence
if and only if $\omega $, introduced 
through (\ref{26}), is $s$-exact modulo 
$d$. By direct computation, we infer 
\begin{eqnarray}
&&\left( S_{1},S_{1}\right) =
f_{\;\;\left[ bc\right. }^{a}f_{\;\;\left.
de\right] }^{c}\int d^{4}x
\left( \eta _{a}^{*}C^{b}C^{e}\eta
^{d}+C_{a}^{*\mu }C^{b}C^{d}
C_{\mu }^{e}-\right. \nonumber \\ \label{47}
&&\left. 2C_{a}^{*\mu }C^{d}
\eta ^{b}A_{\mu }^{e}+\frac{1}{3}%
C_{a}^{*}C^{b}C^{d}C^{e}+
\varepsilon ^{\mu \nu \lambda \rho }A_{\nu
}^{b}A_{\rho }^{e}F_{a}^{\mu \nu }
C^{d}\right) +\nonumber \\
&&2f_{\;\;bc}^{a}f_{ade}
\varepsilon ^{\mu \nu \lambda \rho }\int
d^{4}xA_{\lambda }^{b}A_{\rho }^{c}
A_{\nu }^{e}\partial _{\mu }C^{d}\equiv
\int d^{4}x\omega . 
\end{eqnarray}
From (\ref{47}) we find that $\omega $ 
is $s$-exact modulo $d$ if and only
if the constants $f_{\;\;bc}^{a}$ 
fulfill the Jacobi identity 
\begin{equation}
f_{\;\;\left[ bc\right. }^{a}
f_{\;\;\left. de\right] }^{c}=0.  \label{48}
\end{equation}
Consequently, it follows that 
\begin{equation}
\beta =-\frac{1}{4}f_{\;\;bc}^{a}
f_{ade}\varepsilon ^{\mu \nu \lambda \rho
}A_{\mu }^{b}A_{\nu }^{c}
A_{\lambda }^{d}A_{\rho }^{e},  \label{49}
\end{equation}
which gives the piece of order 
$g^{2}$ from the deformed solution to the
master equation under the form 
\begin{equation}
S_{2}=-\frac{1}{4}f_{\;\;bc}^{a}
f_{ade}\int d^{4}xA_{\mu }^{b}A_{\nu
}^{c}A_{\lambda }^{d}A_{\rho }^{e}.  \label{50}
\end{equation}
By direct computation we find that 
$\left( S_{2},S_{2}\right) =0$, such that
the higher-order equations in the 
deformation parameter are satisfied under
the choice 
\begin{equation}
S_{3}=S_{4}=\cdots =0.  \label{51}
\end{equation}
By virtue of the above results, we 
can conclude that the complete solution
to the master equation (\ref{22}) 
defining our deformation problem is
expressed by 
\begin{eqnarray}
&&\bar{S}=\int d^{4}x\left( -
\frac{1}{4}\varepsilon _{\mu \nu \lambda \rho
}H_{a}^{\mu \nu }H^{a\lambda \rho }+
A_{a}^{*\mu }\left( \left( D_{\mu
}\right) _{\;\;b}^{a}C^{b}+
C_{\mu }^{a}\right) -\frac{1}{2}%
gC_{a}^{*}f_{\;\;bc}^{a}C^{b}C^{c}-
\right. \nonumber \\ \label{52}
&&\left. gC_{a}^{*\mu }f_{\;\;bc}^{a}C^{b}
C_{\mu }^{c}+C_{a}^{*}\eta
^{a}-C_{a}^{*\mu }\left( D_{\mu }
\right) _{\;\;b}^{a}\eta ^{b}-g\eta
_{a}^{*}f_{\;\;bc}^{a}\eta ^{b}C^{c}\right) ,  
\end{eqnarray}
where the notation $\left( D_{\mu }
\right) _{\;\;b}^{a}$ stands for the
covariant derivative 
\begin{equation}
\left( D_{\mu }\right) _{\;\;b}^{a}=
\delta _{\;\;b}^{a}\partial _{\mu
}+gf_{\;\;bc}^{a}A_{\mu }^{c},  \label{53}
\end{equation}
while the deformed field strength 
$H^{a\mu \nu }$ is given by 
\begin{equation}
H^{a\mu \nu }=F^{a\mu \nu }-
gf_{\;\;bc}^{a}A^{\mu b}A^{\nu c}.  \label{54}
\end{equation}
Let us analyze now the deformed 
theory, described by (\ref{52}). We observe
that the antifield-independent 
piece in (\ref{52}) 
\begin{equation}
\bar{S}_{0}=-\frac{1}{4}\int d^{4}x
\varepsilon _{\mu \nu \lambda \rho
}H_{a}^{\mu \nu }H^{a\lambda \rho },  \label{55}
\end{equation}
describes nothing but the topological 
coupling between the vector potentials 
$A_{\mu }^{a}$, known as topological 
Yang-Mills theory. The structure of the
terms linear in the antifields 
$A_{a}^{*\mu }$ shows that our procedure
deforms also the gauge transformations 
\begin{equation}
\bar{\delta}_{\epsilon }A_{\mu }^{a}=
\left( D_{\mu }\right)
_{\;\;b}^{a}\epsilon ^{b}+
\epsilon _{\mu }^{a}.  \label{56}
\end{equation}
Moreover, from the terms $-
\frac{1}{2}gC_{a}^{*}f_{\;%
\;bc}^{a}C^{b}C^{c}-gC_{a}^{*\mu }
f_{\;\;bc}^{a}C^{b}C_{\mu }^{c}$ we learn
that the resulting gauge algebra is deformed, 
while the presence of $%
C_{a}^{*}\eta ^{a}-C_{a}^{*\mu }\left( D_{\mu }
\right) _{\;\;b}^{a}\eta ^{b}$
indicates that the reducibility functions are also deformed 
\begin{equation}
\bar{Z}_{\;\;\alpha _{2}}^{\alpha _{1}}=
\left( \delta _{\;\;b}^{a},-\left(
D_{\mu }\right) _{\;\;b}^{a}\right) .  \label{57}
\end{equation}
In conclusion, the deformation problem 
studied in this paper generates the
deformations of the gauge transformations, 
gauge algebra and reducibility
relations with respect to the starting uncoupled model.

\section{Conclusion}

To conclude with, in this paper we 
have investigated the consistent
interaction that can be introduced 
among a set of  topologically coupled
free vector fields. Starting with the 
BRST differential for the free theory, 
$s=\delta +\gamma $, we initially compute 
the consistent first-order
deformation with the help of some 
cohomological arguments related to the
free model. Next, we prove that the 
deformation is also second-order
consistent and, moreover, matches the 
higher-order deformation equations. As
a result, we are led precisely to the 
topological Yang-Mills theory, that
implies the deformation of the gauge 
transformations, their algebra and of
the accompanying reducibility relations.

\end{document}